\documentclass[12pt]{aastex}
\usepackage{psfig}
\def\degr{\hbox{$^\circ$}}

\def\gsim{\mathrel{\hbox{\rlap{\lower.55ex \hbox {$\sim$}}
                   \kern-.3em \raise.4ex \hbox{$>$}}}}
\def\lsim{\mathrel{\hbox{\rlap{\lower.55ex \hbox {$\sim$}}
                   \kern-.3em \raise.4ex \hbox{$<$}}}}

\begin{document}  
\title{Evolution of spiral shocks in U Gem during outburst}
\author{P.J. Groot\altaffilmark{1}}

\altaffiltext{1}{Harvard-Smithsonian Center for Astrophysics, 60
Garden Street, Cambridge, MA 02138, USA; {\tt pgroot@cfa.harvard.edu}}

\begin{abstract}
Time resolved spectroscopic observations of U Gem during its March
2000 outburst show strong spiral shocks in the accretion disk. 
During the plateau at maximum brightness the spiral shocks contribute
$\sim$14\% to the total He\,{\sc II} flux. The two arms of the spiral show a
distinctly different evolution during the outburst and decline, which
indicates an asymmetric evolution in the disk. 
\end{abstract}


\setcounter{footnote}{0}
\section{Introduction}
One of the major problems in accretion disk physics is the transport
of angular momentum outwards through the disk. Known sources of
viscosity fail by many orders of magnitude, signifying the presence
of one or more additional mechanisms that should be very efficient in
carrying off the angular momentum of disk material. One of these
possible mechanisms is transport of angular momentum through tidally
induced spiral arms.  Although they had been predicted many times
theoretically, first by Sawada, Matsuda and Hachisu in 1986, spiral
shocks have only recently been detected observationally in accretion
disks in Cataclysmic Variables (CVs). They were first detected in IP
Peg in outburst (Steeghs, Harlaftis and Horne, 1997) and have now
been confirmed in the same source (Harlaftis et al., 1999) and in EX
Dra (Joergens, Spruit and Rutten, 2000). IP Peg, as well as EX Dra,
belongs to the dwarf nova class of CVs. These are characterized by 2-5
magnitude outbursts that occur on irregular timescales (ranging from
weeks to decades; see Warner, 1995, for an overview of CVs).  It is
during these outbursts that the spiral arms have been
detected.  

As shown by Steeghs and Stehle (1999; hereafter S\&S99) the low Mach
numbers (which is the ratio of the local gas velocity, often assumed
Keplerian, and the speed of sound in the disk) and the large size of
these disks make the expected spiral pattern an open two-armed
spiral (e.g. Sawada et al., 1986, R\'o\.zycka and Spruit 1993;
Heemskerk 1994). This also causes them to be detected quite readily in
spectroscopic observations.  If spiral arms persist in accretion disks
in quiescence they are expected to be more tightly wound due to the
higher Mach numbers and the smaller size of the disk. This makes them
much harder to be detected in spectroscopic observations (S\&S99). Due
to the very limited sample of systems with detected spiral arms (only
two), and the limited number of observations during their outbursts,
much of the origin, dynamics and evolution of spiral arms in accretion
disks remains a mystery. Their dependence on outburst phase, orbital
period, mass-ratio's etc. is still largely unknown. Clearly a larger
sample is needed. In this {\sl Letter} I report on observations taken
of U Gem, before and during its outburst in March 2000.

U Gem is the prototypical and first discovered CV (Hind, 1855). It is
a dwarf nova system with a relatively long orbital period of 4h17m. It
has been studied extensively photometrically (see e.g. Smak, 1993, and
references therein) and shows grazing eclipses: the accretion disk is
partially eclipsed, but the white dwarf remains visible during
mid-eclipse.  From photometry the system parameters have been deduced
by many authors, but I will use the orbital period and ephemeris as
determined by Marsh et al. (1990) and the values of Friend et
al. (1990) for the mass of the white dwarf primary, $M_1$ (1.24
M$_{\odot}$), the mass ratio, $q$ (=$M_2$/$M_1$=0.46) and the
inclination $i$ (69.7\degr). These give the projected velocity
amplitudes $K_1$ and $K_2$ that are needed for the Doppler mapping
procedure.

U Gem is known to undergo outbursts on timescales between 30 and 250
days (Warner 1995). The outburst in the spring of 2000 started on
March 1, and lasted through March 21. In Figure 1 I show the
photometric light curve of this outburst, compiled from observations
of the American Association of Variable Star Observers (AAVSO; Mattei,
2000).

\section{Observations and Data reduction}

Spectroscopic data was obtained by Mike Calkins and Perry Berlind on
eight different nights, using the 1.5m Tillinghast Telescope at the
Fred Lawrence Whipple Observatory on Mt. Hopkins, AZ. The high
throughput FAST spectrograph (Fabricant et al., 1998) was equipped
with a Loral 512x2688 pixel CCD. We used a 1200 line grating, centered
on 5050 \AA, to cover 1000 \AA\ at a resolution of 1.1\AA. For the
wavelength calibration arc spectra were taken every 30 minutes. An
overview of the observations is given in Table 1.

The data was reduced using the ESO-MIDAS package with additionally written
software. All spectra were optimally extracted and wavelength
calibrated after standard de-biasing and flatfielding.

Although the spectrum covers multiple lines, I will in further concentrate
on the spectral evolution of He\,{\sc ii} 4686 \AA, since this line most
clearly shows the spiral arms during outburst. A full discussion of
the spectral evolution of  U Gem during the outburst will be given
elsewhere. 

\section{Spectral evolution of He\,{\sc ii} 4686\AA.}

\begin{figure}
\centerline{\psfig{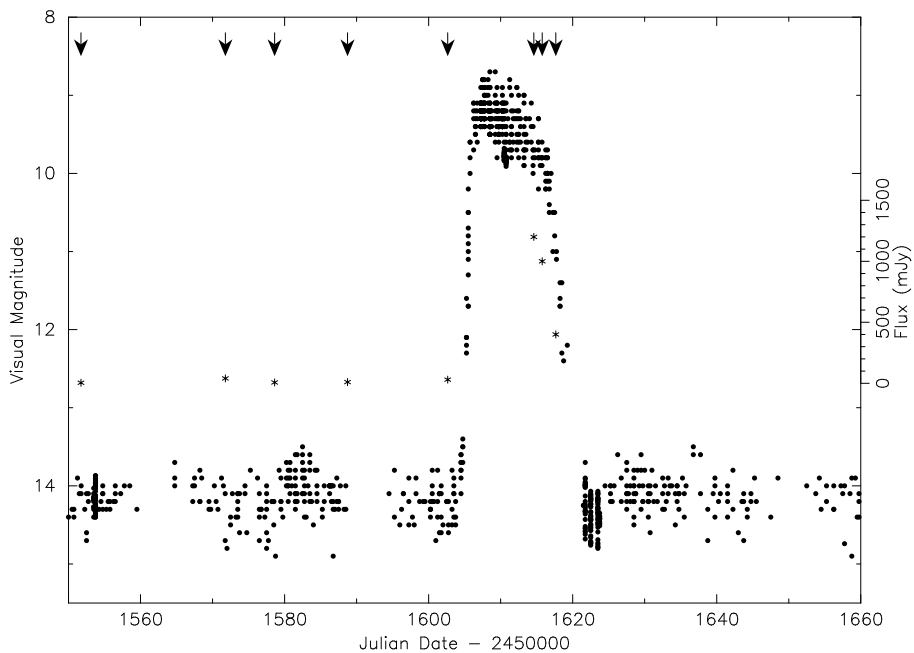}}
\caption[]{The outburst light curve as observed by members of the
AAVSO (Mattei; 2000). Arrows indicate the times of our spectroscopic
observations. Stars indicate the approximate continuum level (on the
scale on the right) of the spectrosopic data. \label{fig:aavso}}

\end{figure}

Based on the outburst light curve of U Gem shown in Fig.\
\ref{fig:aavso}, the observations are divided into four episodes
(see Table 1): the observations from JD 2451551 until JD 245602 when
the system was in quiescence form Episode I, the outburst 
observations at JD 2451614 constitute Episode II, at JD 2451615
constitute Episode III and at JD 2451617 constitute Episode IV.
 
\begin{figure*}
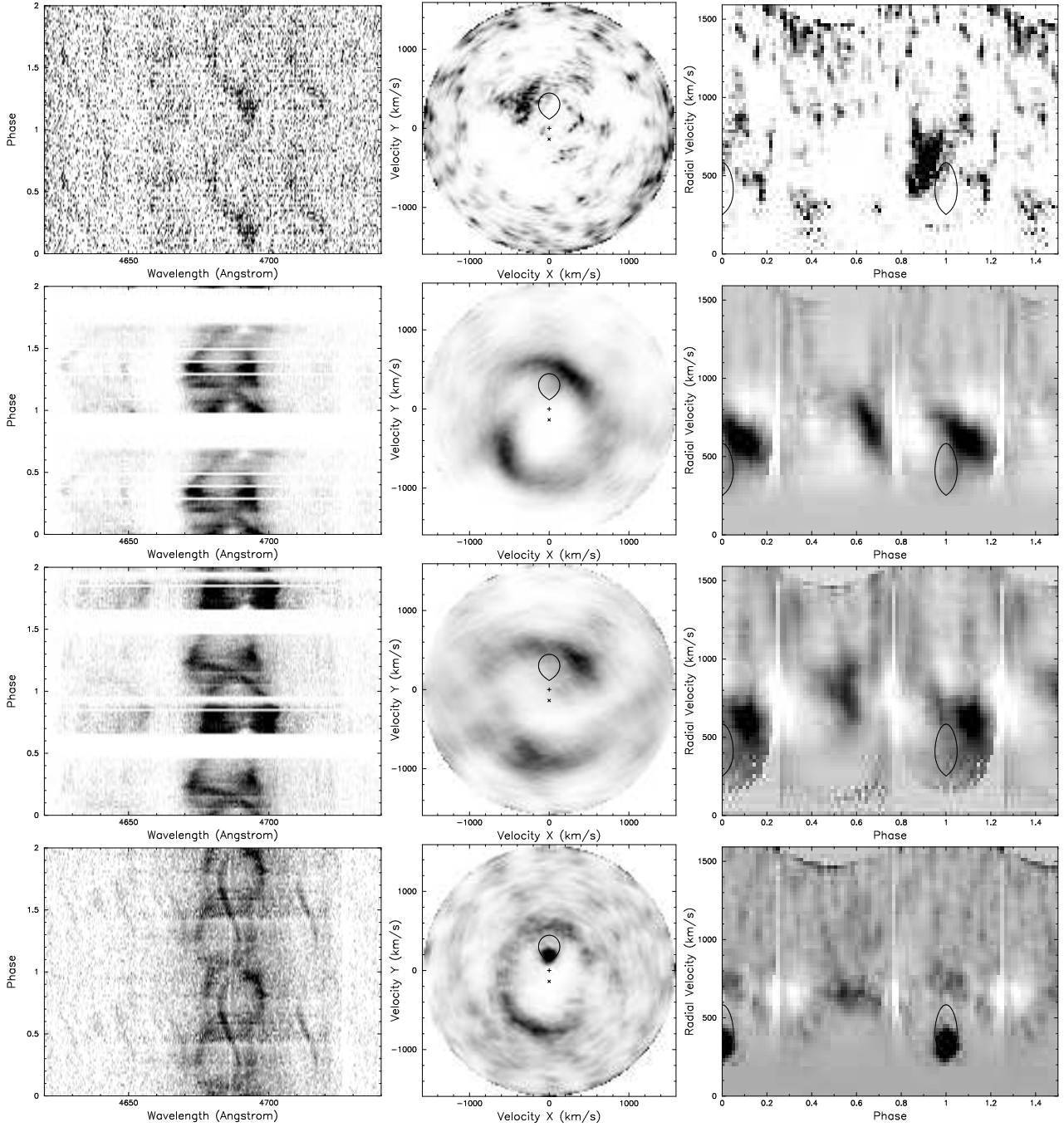


\begin{minipage}{6cm}
\centerline{\psfig{figure=trailHeII_1.ps,width=6cm,angle=-90}}
\end{minipage}
\begin{minipage}{4.65cm}
\centerline{\psfig{figure=dopmapHeII_1.ps,width=4.6cm,angle=-90}}
\end{minipage}
\begin{minipage}{6cm}
\centerline{\psfig{figure=maprphiHeII_1.ps,width=6cm,angle=-90}}
\end{minipage}

\begin{minipage}{6cm}
\centerline{\psfig{figure=trailHeII_3.ps,width=6cm,angle=-90}}
\end{minipage}
\begin{minipage}{4.65cm}
\centerline{\psfig{figure=dopmapHeII_3.ps,width=4.6cm,angle=-90}}
\end{minipage}
\begin{minipage}{6cm}
\centerline{\psfig{figure=maprphiHeII_3.ps,width=6cm,angle=-90}}
\end{minipage}

\begin{minipage}{6cm}
\centerline{\psfig{figure=trailHeII_4.ps,width=6cm,angle=-90}}
\end{minipage}
\begin{minipage}{4.65cm}
\centerline{\psfig{figure=dopmapHeII_4.ps,width=4.6cm,angle=-90}}
\end{minipage}
\begin{minipage}{6cm}
\centerline{\psfig{figure=maprphiHeII_4.ps,width=6cm,angle=-90}}
\end{minipage}

\begin{minipage}{6cm}
\centerline{\psfig{figure=trailHeII_5.ps,width=6cm,angle=-90}}
\end{minipage}
\begin{minipage}{4.65cm}
\centerline{\psfig{figure=dopmapHeII_5.ps,width=4.6cm,angle=-90}}
\end{minipage}
\begin{minipage}{6cm}
\centerline{\psfig{figure=maprphiHeII_5.ps,width=6cm,angle=-90}}
\end{minipage}

\caption[]{ The Episode I - IV (top to bottom) data on He\,{\sc ii}
4686: the trailed spectra (left), the Doppler maps (middle) and the
$v_{\rm r}\phi$ maps (right). In the Doppler tomograms the balloon
represents the secondary, the `+' sign the center of mass, the `x' the
position of the white dwarf. In the $v_{\rm r}\varphi$-map the balloon
indicates the position of the secondary. Data at $\varphi>$1 is copied
from the same phases at $\varphi<$1 for display purposes.
\label{fig:heII}}
\end{figure*}

Figure\ \ref{fig:heII} shows the trailed spectra of He\,{\sc ii}
$\lambda$4686 (and during quiescence He\,{\sc i} $\lambda$4713 and
during outburst the Bowen blend of C\,{\sc iii}/N\,{\sc iii} at $\lambda$4650)
in the left panel for the four different episodes. In the middle panel
of Fig.\ \ref{fig:heII} I show the Doppler maps (Marsh and Horne,
1988) of the trailed spectra on the left. Doppler tomography gives a
time-averaged (over one or more orbits) remapping of the velocity
information in the trailed spectra into velocity space. In the process
of the Doppler mapping the azimuthal structure of the system is
preserved, but the radial structure is inverted in a non-linear
way. Since the (partial) eclipse violates the Doppler mapping
assumption of equal visibility at all phases, the phases between 0.88
and 0.12 have not been included in the Doppler mapping.

The right panel of Fig.\ \ref{fig:heII} shows a transformation of the
Doppler maps, the $v_{\rm r}\varphi$-plot, in which radial velocity
from the white dwarf is shown as a function of the orbital phase.
Here I retained the normal photometric phase. In this definition the
secondary is located at phase zero. This representation is basically
the two dimensional version of Fig. 4 in Harlaftis et al., (1999).
The $v_{\rm r}\varphi$-plots shown here have the symmetric part of the
disk emission subtracted. This symmetric part is determined by taking
the median over all phases at each radial velocity. In this way we can
estimate the fraction of flux coming from the spiral arms, although
this will be a slight underestimate since the spiral arms are included
in the determination of the median. The advantage of this projection
is that the spiral arms map into linear features, whose relative
positions and slopes can be determined more accurately than in a
conventional Doppler map.

\subsection{Pre-outburst}

During Episode I the trailed spectrum shows a clear single
S-wave. Doppler tomography identifies the hot-spot to be the origin of
this He\,{\sc ii} emission. The velocities lie in between the accretion
stream trajectory and the Kepler velocity trail. This is a normal
situation for hot-spot emission, where material in the disk and in the
accretion stream mixes. From the position of the He\,{\sc ii} emission we derive
that the outer edge of the hot-spot is located at a position of 0.55
R$_{\rm L_1}$ and is located at an angle 20\degr\ away from the line
joining the centers of the two stars, very similar to the results
obtained by Marsh et al. (1990). The hot spot emission accounts
for 46\% of the total He\,{\sc ii} flux at this episode. 

\subsection{Outburst: Accretion disk and secondary star emission}

During the outburst episodes II, III and IV, clear spiral shocks are
present in the disk in U Gem. As we can see from Fig.\
\ref{fig:aavso}, Episode II and III were taken at the end of the
maximum brightness plateau and Episode IV is taken during the decline
of the outburst. During the outburst the non-symmetric part of the
emission produced 14\% (Episode II), 5.5\% (Episode III) and 9.5\%
(Episode IV) of the total He\,{\sc ii} flux. In Episode IV the
majority of the 9.5\% (7.8\%) is on account of emission from the
secondary. This secondary star emission is completely absent in
Episodes II and III, showing that at this stage of the outburst the
secondary, but not the spiral shock, is completely blocked from
irradiation by the high energy radiation from the inner disk. 

\subsection{Outburst: Spiral arm evolution}

Comparing the position and strength of the spiral arms between
Episodes II, III and IV, we see not only a clear evolution of the
spiral arms, but also a clear difference between the two arms. Let us
label the two spiral arms as S0.6 for the one that is located around
phase 0.6 and S0.1 for the one that is visible at phase 0.1 (the
`lower' and `upper' arm, respectively, in the Doppler maps in Fig.\
\ref{fig:heII}). We can see in the right hand panel of Fig.\
\ref{fig:heII} that S0.1 is stronger than S0.6 in Episode II and III,
but not in Episode IV. However, the more remarkable differences are in the
slopes of the spiral arms and the changes in these slopes. In Episode
II the slope of S0.6 is --2600$\pm$700 (estimated 3$\sigma$)
\mbox{km\,s$^{-1}$\,orbit$^{-1}$} and for S0.1 this is --1300$\pm$200
\mbox{km\,s$^{-1}$\,orbit$^{-1}$}. In Episode III these numbers are
--1740$\pm$400 \mbox{km\,s$^{-1}$\,orbit$^{-1}$} for S0.6 and
--1100$\pm$200 \mbox{km\,s$^{-1}$\,orbit$^{-1}$} for S0.1 and in
Episode IV --250$\pm$140 \mbox{km\,s$^{-1}$\,orbit$^{-1}$} for S0.6
and --900$\pm$170 \mbox{km\,s$^{-1}$\,orbit$^{-1}$} for S0.1. We see
that, although for both arms the slopes decrease, for S0.6 this
decrease is much more rapid and pronounced than for S0.1. 

Judging the extent of the spiral arms is much more difficult and
highly subjective. From the $v_{\rm r}\varphi$ plots I estimate the
phase extent of S0.6 to be 0.5-0.8 (Episode II), 0.4-0.6 (Episode III)
and 0.4-0.7 (Episode IV) and for S0.1: 0.9-0.3 (Episode II), 0.9-0.2
(Episode III) and 0.9-0.4 (Episode IV), although the upper limit on
the last number is arguable, and could also be 0.2, depending on what
one believes is still genuine spiral arm emission.

\section{Comparison with IP Peg, EX Dra and simulations}

Spiral shocks have been convincingly detected in the
outburst spectra of IP Peg (Steeghs et al., 1997 and Harlaftis et al.,
1999) and EX Dra (Joergens et al., 1999). Especially the comparison
with IP Peg is of relevance since this source has been observed during
rise and maximum of an outburst, complementing the end of maximum and
decline data shown here. 

Comparing our Episode II data with the outburst maximum maps shown in
Harlaftis et al. (1999) shows that there are similarities and clear
differences. The fraction of He\,{\sc ii} flux in the spiral arms
($\sim$15\%) is the same in both systems, but in U Gem there is no
secondary star emission at all at the maximum of the outburst, whereas
this is clearly seen in IP Peg. Also the spiral arms in U Gem seem to
be rotated in phase towards later phases. The outburst maximum
tomogram in IP Peg more closely resembles our Episode IV tomogram,
which was taken during the decline of the outburst.  The same
comparison holds true for the spirals found in EX Dra, also taken at
outburst maximum.

Comparison with simulations, most notably those of S\&S99, 
shows a striking resemblance between the Episode II tomogram
of U Gem presented here, and the low Mach number simulations shown
there (e.g. their Figure 8). This is in contrast with the IP Peg, EX
Dra and U Gem Episode IV tomograms, which all appear to have their
maximum emissivity rotated anti-clock wise with respect to the
simulations of S\&S99. 

\section{Discussion}

A possible evolutionary scenario for the spiral shocks one can derive
from the comparison of these three systems is that the shocks appear
immediately when the outburst starts (IP Peg; Steeghs et al., 1997),
but not just prior to outburst (our Episode I), grow in strength when
the outburst reaches its maximum magnitude (Harlaftis et al., 1999 and
Joergens et al., 2000), and continue to gain in strength, or at least
remain constant, during the plateau phase characteristic of many DN
outbursts (our Episodes II and III), and then fade  
during outburst decline (our Episode IV).

This interpretation, however, neglects any differences in system
parameters between U Gem, EX Dra and IP Peg. For instance, of these
systems U Gem has the lowest inclination (69\degr\ vs. 80\degr\ for IP
Peg and 84\degr\ for EX Dra), the most massive white dwarf
(1.24M$_{\odot}$, vs. 1.0 M$_{\odot}$ for IP Peg and 0.75 M$_{\odot}$
for EX Dra) and the most extreme mass ratio, $q$=$M_2/M_1$=0.46,
vs. 0.64 for IP Peg and 0.74 for EX Dra (system parameters of IP Peg
from Friend et al., 1990 and for EX Dra from Fiedler et al., 1997). All of
these factors may be of (unknown relative) importance for the
visibility and strength of the spiral shocks during outburst. 

One of the main questions in spiral shock research is the question
whether the shocks persist through quiescence and could serve as a
funnel to transport the angular momentum outwards. So far, no spiral
arms have been detected in quiescent dwarf novae. The data presented
here show that the shocks certainly fade during the decline of the
outburst. However, it is not clear if they also wrap up at the same
time. S\&S99 have shown that the spiral arms in quiescence should be
tightly wrapped and will be difficult to observe. The rapid decrease
in spiral shock slope (especially of the S0.6 arm) in the $v_{\rm r}\varphi$
diagram could be taken as an indication of a 'wrapping' up of the
spiral arms, which would, to the observer, show up as a
'circulization', i.e. a flattening of the slope, of the spiral
shocks. The data presented here is, however, too scarce to draw a
strong conclusion on this.

To the best of my knowledge, no theoretical investigation or modelling
has shown a difference in the evolution of the spiral shock between
the two arms, which is clearly seen in U Gem. 

To disentangle the effect of spiral shock evolution and differences in
system parameters, outbursting dwarf novae (at least IP Peg, EX Dra
and U Gem) should be followed spectroscopically during a complete
outburst. Certainly for U Gem, which reaches V$\sim$9 during outburst
maximum, this should not be a difficult task.

\vspace{1cm} {\bf Acknowledgments} I wish to thank Mike Calkins and
Perry Berlind for obtaining the observations and Henk Spruit for the
use of his {\sc dopmap} programs. In this research, I have used, and
acknowledge with thanks, data from the AAVSO International Database,
based on observations submitted to the AAVSO by variable star
observers worldwide. I thank the referee for her/his comments which
improved the paper.  PJG is supported by a Harvard-Smithsonian CfA
fellowship.

\references
Fabricant, D., Cheimets, P., Caldwell, N, and Geary, J., 1998, PASP
110, 79\\
Fiedler, H., Barwig, H. and Mantel, K.H., 1997, A\&A 327, 173\\ 
Friend, M.T., Martin, J.S., Smith, R.C., Jones, D.H.P.,1990, MNRAS 246, 637\\
Groot, P.J., 2001, {\sl in preparation}\\
Harlaftis, E.T., Steeghs, D., Horne, K., Mart\'{\i}n, E., and Magazzu,
A., 1999, MNRAS, 306, 348\\
Heemskerk, M., 1994, A\&A 288, 807\\
Hind, J.R., 1855, MNRAS 16, 56\\
Joergens, V., Spruit, H.C. and Rutten, R.G.M., 2000, A\&A 356, L33\\
Marsh, T.R and Horne, K., 1988, MNRAS 235, 269\\
Marsh, T.R., Horne, K., Schlegel, E.M., Honeycutt, R.K., Kaitchuck,
R.H. 1990, ApJ 364, 637\\
Mattei, J., 2000, {\sl private communications}\\
R\'o\.zycka, M. and Spruit, H.C., 1993, Apj 417, 667\\
Sawada, K., Matsuda, T. and Hachisu, I., 1986, MNRAS 219, 75\\
Smak, J., 1993, AA 43, 121\\
Steeghs, D., Harlaftis, E.T. and Horne, K., 1997, MNRAS 290, L28\\
Steeghs, D., and Stehle, R., 1999, MNRAS 307, 99\\
Warner, B., 1995, Cataclysmic Variable Stars, Cambridge Astrophysics
Series 28, Cambridge Univ. Press, Cambridge, UK.

\begin{table}
\caption[]{Log of observations of U Gem as observed with the 1.5m
Tillinghast Telescope. Each observation was 270s. See text for the
explanation of the division in four episodes.}
\begin{minipage}{8cm}
\begin{tabular}{lllll}
\hline
Date & HJD Start--2450000 & Phase coverage & No. Obs. & Episode \\
07/01/2000 &  1551.72567 & 78645.65 - 78647.50 & 95 & I\\
27/01/2000 &  1571.78126 & 78759.01 - 78760.07 & 54 & I \\
03/02/2000 &  1578.61545 & 78798.71 - 78799.75 & 108& I \\
13/02/2000 &  1588.74123 & 78854.88 - 78856.06 & 60 & I  \\
27/02/2000 &  1602.65477 & 78933.53 - 78934.58 & 54 & I \\
10/03/2000 &  1614.61133 & 79001.98 - 79002.68 & 75 & II \\
11/03/2000 &  1615.76881 & 79007.66 - 79008.44 & 40 & III\\
13/03/2000 &  1617.65608 & 79018.33 - 79019.38 & 54 & IV \\

\end{tabular}
\end{minipage}
\end{table}

\end{document}